\documentstyle[12pt,aasms4]{article}
\begin{document}

\title{Quasars around the Seyfert Galaxy NGC3516}
\author{Yaoquan Chu}
\affil{Center for Astrophysics, University of Science and Technology of
China, Hefei, Anhui 230026, China}

\author{Jianyan Wei}
\affil{Beijing Astronomical Observatory, Chinese Academy of Science, Beijing
100080, China}

\author{Jingyao Hu}
\affil{Beijing Astronomical Observatory, Chinese Academy of Science, Beijing
100080, China}

\author{Xingfen Zhu}
\affil{Center for Astrophysics, University of Science and Technology of
China, Hefei, Anhui 230026, China}

\author{H. Arp}
\affil{Max-Planck-Institut f\"ur Astrophysik, Karl-Schwarzschild-Str.\ 1,
85740 Garching, Germany}

\begin{abstract}
We report redshift measurements of 5 X-ray emitting blue
stellar objects (BSO's) located less than 12 arc min from the X-ray
Seyfert galaxy, NGC 3516. We find these quasars to be distributed
along the minor axis of the galaxy and to show a very good correlation
between their redshift and their angular distance from NGC
3516. Moreover the redshifts of these 5 quasars are: 0.33, 0.69,
0.93, 1.40 and 2.10 which are very near the peaks of the redshift
periodicity distribution (i.e.\ $z=0.3$, 0.6, 0.96, 1.41 and 1.96). All
these observed properties strikingly confirm, around this single
example of a Seyfert, the composite picture derived from previous
physical associations of quasars with low redshift, active galaxies. 
\end{abstract}

\keywords{quasars: general -- Seyfert galaxies}

Recently H.-D. Radecke (1997) has presented evidence that there is a
significant (minimum of 7.4 sigma) excess of bright X-ray sources
around a nearly complete sample of Seyfert galaxies. Arp (1997) had
inspected each individual X-ray map of 24 Seyfert galaxies with
an apparent magnitude between $8.04 \le B_T^{0,i} \le 12.90$ mag.\ and
found that on Schmidt Survey plates most of these excess X-ray sources
are identified with blue stellar objects (BSO). These X-ray emitting
BSO's generally show pairing and alignment configurations cross the
central Seyfert galaxies. 

We have started a program to obtain the optical spectra of these X-ray
emitting BSO's. Herewith we present our result in the field
NGC3516. NGC3516 is a strong X-ray Seyfert galaxy, with apparent
magnitude V=12.40 and redshift z=0.009. In Radecke's Seyfert sample,
NGC3516 belongs to the faintest and most distant category. X-ray
sources associated with this galaxy would therefore be as bright as
$6\times 10^{40}$ erg s$^{-1}$, which is about $\sim 300$ times
brighter than any galactic X-ray sources. There are 5 BSO objects
listed in Table 3 of Arp's paper, these objects are located in the
region $\theta<12^\prime$ and aligned NW--SE across NGC3516. (see
Fig.1 in which these BSO are identified by their X-ray count rate C =
cts$\cdot$ks$^{-1}$ value).\placefigure{fig1}

We obtained the spectra of the 5 BSO's on April 4--5, 1997 using the
2.16m telescope at Xinlong Station, Beijing Astronomical
Observatory. An OMR spectrograph was attached at the Cassegrain focus. A
TEK1024 $\times$ 1024 CCD served as detector at a resolution of
400A/mm. The coverage of the spectra was from 3700 to
8000A. Integration time for each spectrum was 3600s. Fe/Ar lamp was
used for wavelength calibration and Feige34 and Hz44 as flux standard
stars. The data were reduced with the IRAF package. One of the objects
Q1107+7232 (C=7.1) is already listed in the Hewitt-Burbidge quasar
catalog (1993). It has redshift z = 2.10. We found the other four
objects all to be quasars. Their redshifts are listed in the following
Table 1.

We would like to thank E. Margaret Burbidge for allowing us to quote her measures with the 3 meter 
reflector at Lick Observatory. For the last two quasars in Table~1 she reports independently
measured redshift values of $z=0.68$ and 0.33 (private communication).

\section{Decreasing redshift as the quasars increase their distance
from the galaxy}

It is of interest to note that for these 5 quasars we find there is a
very good linear correlation between redshift z and $\theta$, the
angular distance from the center of NGC 3516.

The statistical analyses show that the linear regression is z=3.06 --
0.22$\theta$, with the correlation coefficient $=-0.957$ and the
standard error of the regression line (Sy/x) = 0.23. We note also that along
the NW--SE alignment of these quasars, at $\theta \sim
22.5^\prime$ there is a very strong X-ray source which is listed as having a
Seyfert spectrum (V\'eron and V\'eron 1996) with redshift z = 0.089
(about 10 times the redshift of NGC3516). Optically it is a compact,
semi-stellar object. With its strong X-ray and radio properties it is
closely allied to BL Lac objects and therefore to the transition
between quasars and objects with increasing amount of stellar population.
When we consider these objects together, in Fig.2 we plot redshift z
against the natural logarithm of $\theta$, and find there is a good
correlation: z = 3.86 -- 1.28ln $\theta$. The correlation coefficient
$=-0.942$ and the standard error of the regression line is 0.276. If
these quasars were ejected from the central galaxy, it means that the
younger the quasar, the closer it is to the center, and the higher the
redshift.\placefigure{fig2}

\section{Alignment along galaxy minor axis}
It is shown in Fig.3 that these 5 quasars and the X-ray Seyfert galaxy
in the center of the field lie within $\pm 20$ degree of a line which
is the minor axis of NGC3516. Just the chance that the above 6 objects
could accidentally lie within $\pm 20^{\rm o}$ of a line through the center
of the galaxy is, by itself, only $10^{-4}$. But it is now clear
that Seyfert galaxies eject quasars preferentially along their minor
axes (see Arp 1997b,c). The NGC3516 line of quasars turns out to lie within a few degrees of its minor axis. Therefore the chance of accidental occurrence is about another two orders of magnitude less.\placefigure{fig3}

\section{Quantization of redshifts}
An especially significant result for these six objects is their
specific values of redshifts. It has been known for a long time that
the redshifts of quasars have preferred values at z = 0.06, 0.30, 0.60,
0.96, 1.41 and 1.96 and that these redshift peaks fit the formula
$\Delta \ln (1+z) = 0.206$. 

The fit to the formula is given by Karlsson (1977). Some claims that
the periodicity was a selection effect of redshifting emission lines
through UBV filters has been refuted by showing the effect is not large enough to
be significant and, more directly, by finding the same strong
periodicity in quasars which had been identified from their radio
emission (Arp et al 1990). It has also been shown that modern detectors give full wavelength
coverage of the spectra without gaps (Burbidge 1978). 
And, of course, the present measures in
NGC3516 represent a complete sample of the brightest candidates which
have been selected by X-rays.

As was pointed out by Zhu and
Chu (1990), this periodicity in redshift is clearer for
multiple quasars which are associated with low redshift
galaxies. Here we find that the 5 aligned quasars plus the quasarlike Seyfert have redshift values
 very close to these peaks:

\begin{tabular}{lll}
S1102+7246& z=0.09&$\to$ 0.06\\
Q1108+7226& z=0.33& $\to$ 0.3\\
Q1106+7244& z=0.69& $\to$ 0.6\\
Q1105+7242& z=0.93& $\to$ 0.96\\
Q1105+7238& z=1.40& $\to$ 1.41\\
Q1107+7232& z=2.10& $\to$ 1.96
\end{tabular}

Considering that quasars in general usually define broader peaks, the
quasars in this field are a particularly strong confirmation of this
property which is so difficult to interpret as a Doppler induced redshift.

\section{Summary}
All of the properties of the high redshift X-ray objects in the NGC~3516 field
confirm the body of earlier results on quasars associated with active galaxies.
We conclude that because of the number of objects in this one group, the evidence
has been greatly strengthened that quasars are ejected from nearby active galaxies
and exhibit intrinsic redshifts.

\newpage 

\figcaption[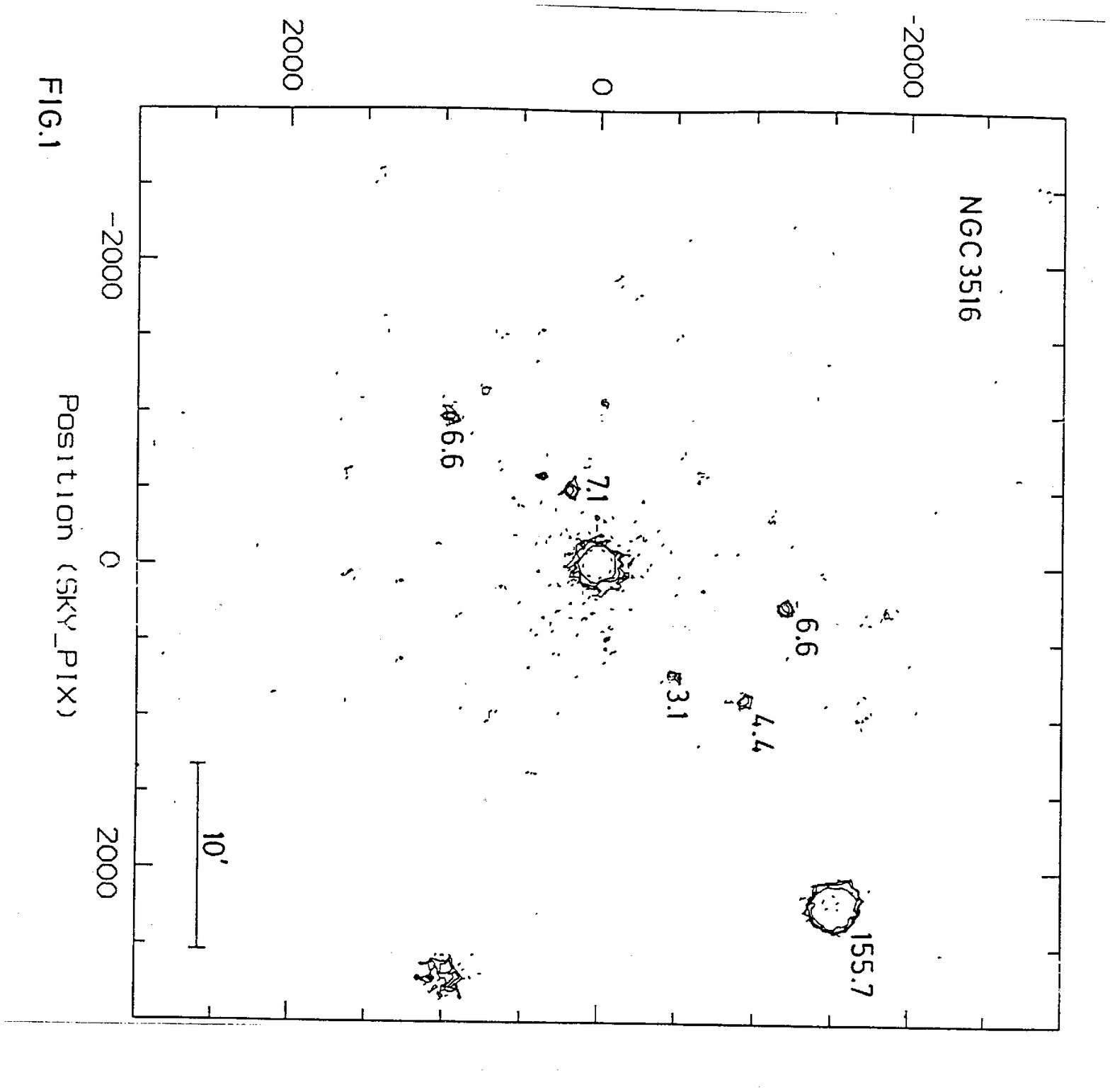]{The ROSAT PSPC map of NGC3516. The five quasars are
identified by their X-ray count rate C (C=cts$\cdot$ks in 0.5--2.0 Kev
band). Coordinate scales are in pixels = 1/2 arc sec.\label{fig1}}

\figcaption[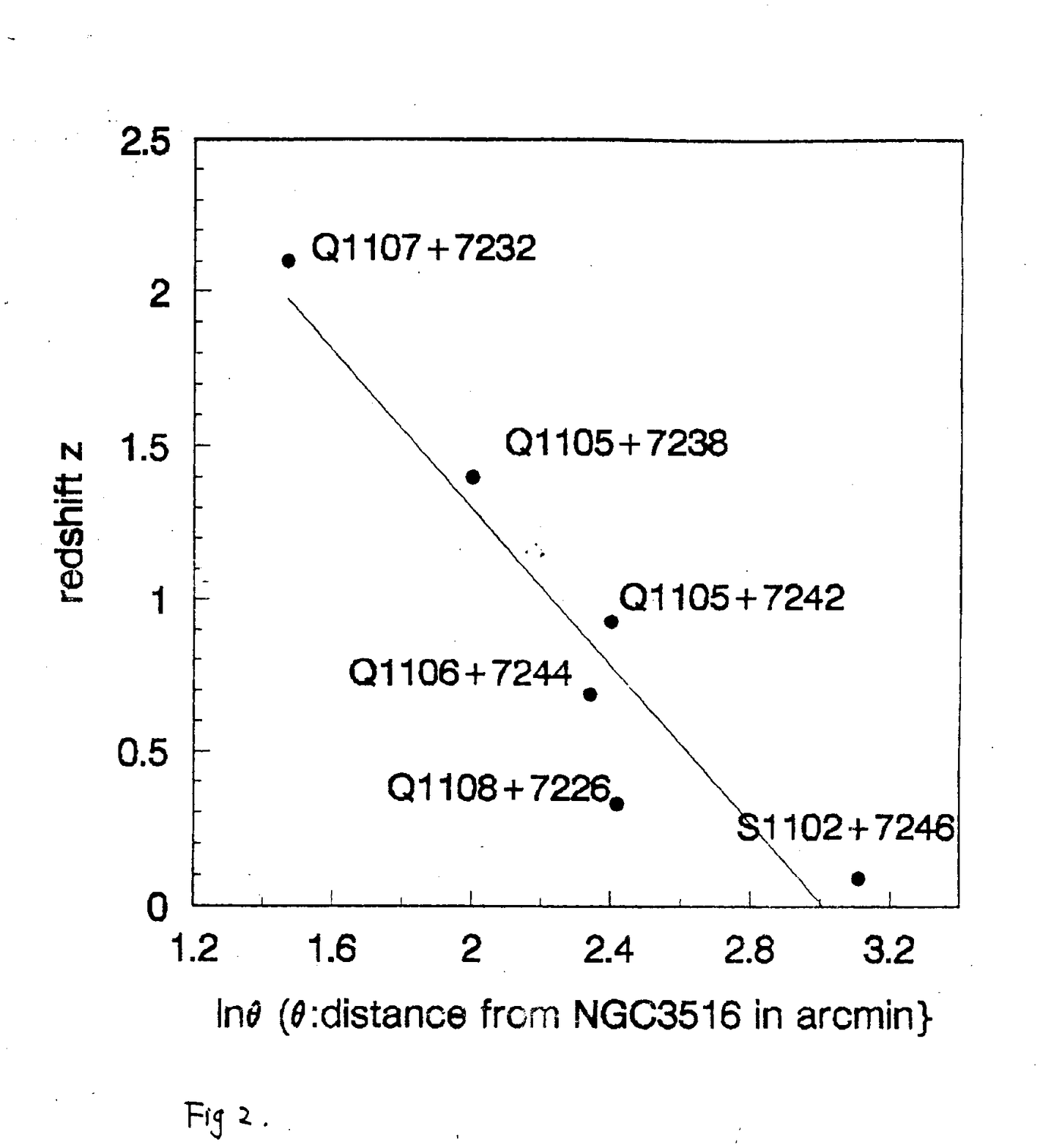]{The redshifts z plotted against natural logarithms of $\theta$, the
angular distance from central galaxy NGC3516. The line represents the
linear regression.\label{fig2}}

\figcaption[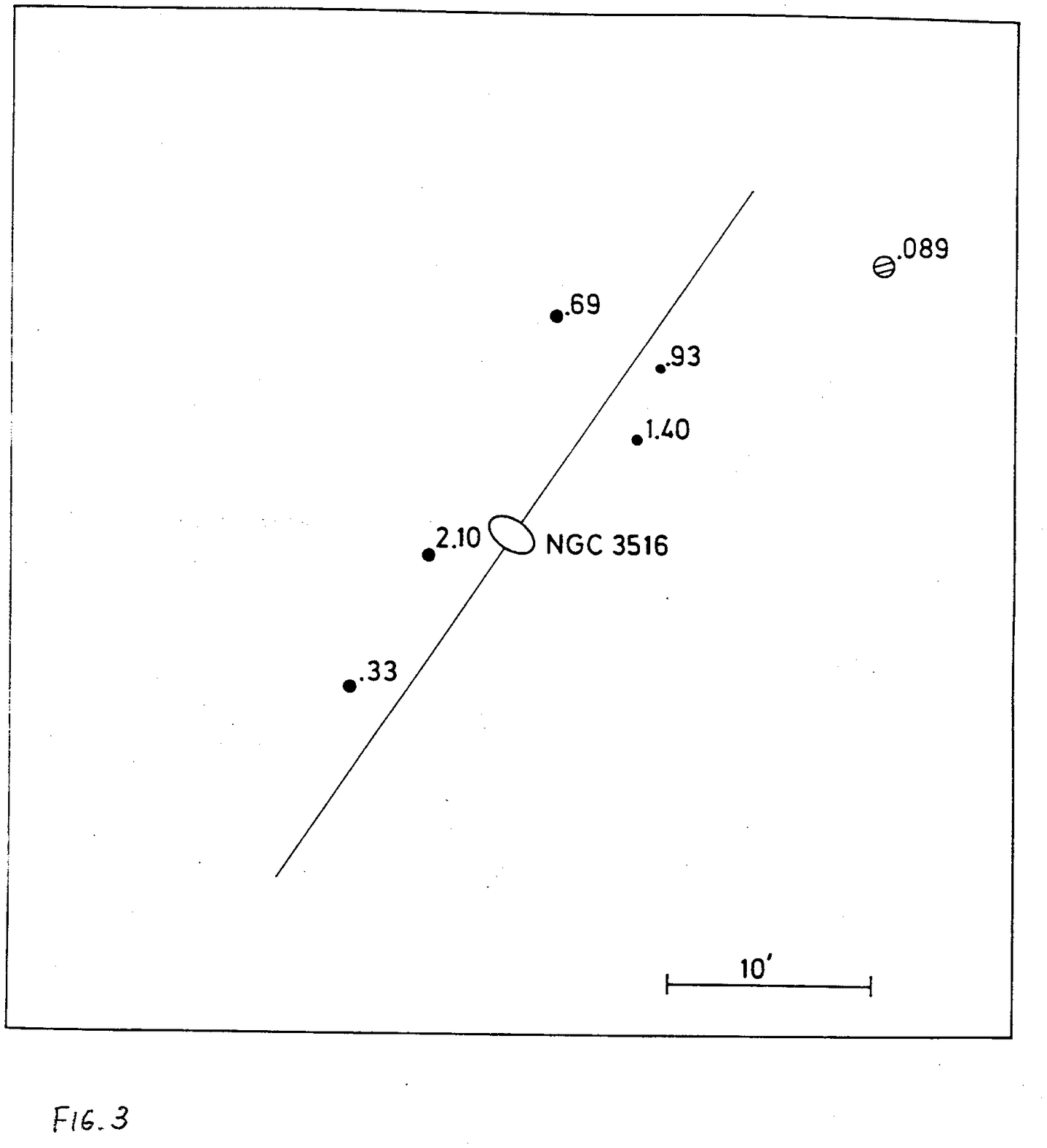]{The six objects which are brightest in X-rays in the NGC3516
field. Redshifts are written to the upper right of each object. The
redshift of NGC 3516 is z=0.009.\label{fig3}}

\newpage

\label{table1}
\begin{deluxetable}{rrrr}
\footnotesize
\tablehead{
\colhead{QSO}&\colhead{Line}&\colhead{$\lambda_{obs}$}&\colhead{z}}
\startdata
Q1105+7242&CIII]1909&3658&0.916\nl
(C=4.4)&MgII 2798&5428&0.940\nl
&$H_\gamma$ 4340&8391&0.932\nl
\tableline
&&&mean z=0.929\nl\\

Q1105+7238&CIV 1549&3727&1.406\nl
(C=3.1)&CIII]1909&4569&1.393\nl
&MgII 2798&6706&1.397\nl
\tableline
&&&mean z=1.399\nl\\

Q1106+7244&MgII 2798&4729&0.690\nl
(C=6.6N)&$H_\delta$ 4102&6944&0.693\nl
&$H_\gamma$ 4340& 7324& 0.688\nl
&$H_\beta$ 4861& 8210& 0.689\nl
\tableline
&&&mean z=0.690\nl\\

Q1108+7226&MgII 2798&3702& 0.323\nl
(C=6.6S)& $H_\beta$ 4861& 6468&0.330\nl
&[OIII] 5007&6659&0.329\nl
&$H_\alpha$ 6563& 8734&0.331\nl
\tableline
&&&mean z=0.328\nl

\enddata
\end{deluxetable}

\newpage

\label{table2}
\begin{deluxetable}{lccccc}
\footnotesize
\tablehead{
\colhead{Object}&\colhead{Redshift}&\colhead{Distance from NGC3516}\nl
\colhead{}&\colhead{z}&\colhead{$\theta$ (arcmin)}}
\startdata
Q 1107+7232& 2.1& 4.34\nl
Q 1105+7238& 1.4& 7.42\nl
Q 1105+7242& 0.93& 10.99\nl
Q 1106+7244& 0.69& 10.37\nl
Q 1108+7226& 0.33& 11.23
\enddata
\end{deluxetable}

\end{document}